\documentclass[journal]{IEEEtran}
\usepackage{cite,graphicx}
\usepackage{amsmath,amssymb,amsthm}
\usepackage{mathtools}
\usepackage{fancyhdr}
\usepackage{mdwmath}
\usepackage{mdwtab}
\usepackage{balance}
\usepackage{xcolor}
\usepackage{bm}
\usepackage{array}
\usepackage{subfig}
\usepackage{amsmath}
\usepackage[ruled,linesnumbered]{algorithm2e}
\usepackage{algorithmic}
\usepackage{multirow}
\usepackage{flafter}
\usepackage{mathrsfs}
\usepackage{url}
\usepackage{multicol}
\usepackage{hyperref}
\usepackage{hhline}
\usepackage{makecell}
\usepackage{tcolorbox}
\tcbuselibrary{breakable}
\usepackage{pdfpages}
\usepackage{enumitem}
\newcommand{\mb}{\mathbf}

\newtheorem{theorem}{Theorem}
\newtheorem{lemma}{Lemma}
\newtheorem{proposition}{Proposition}

\newtheorem{definition}{Definition}
\newtheorem{assumption}{Assumption}

\newcommand{\revisioncolor}{black}

\usepackage[left= 0.625in, right= 0.625in,top=0.76in, bottom=0.97in]{geometry}

\title{Multi-Agent Reinforcement Learning Counteracts Delayed CSI in Multi-Satellite Systems}

\author{\IEEEauthorblockN{
Marios Aristodemou\IEEEauthorrefmark{4},
Yasaman Omid\IEEEauthorrefmark{1}, 
Sangarapillai Lambotharan\IEEEauthorrefmark{2}, Mahsa Derakhshani\IEEEauthorrefmark{1}, Lajos Hanzo \IEEEauthorrefmark{3}
}
\IEEEauthorblockA{\IEEEauthorrefmark{4}Department of Computer Science, University of York, York, U.K.}

\IEEEauthorblockA{\IEEEauthorrefmark{1}Wolfson School of Mechanical, Electrical and Manufacturing Engineering, Loughborough University, Loughborough, U.K.}

\IEEEauthorblockA{\IEEEauthorrefmark{2}Institute for Digital Technologies, Loughborough University (London Campus), London, U.K.}

\IEEEauthorblockA{\IEEEauthorrefmark{3}School of Electronics and Computer Science, University of Southampton, Southampton, U.K.}

\thanks{Copyright (c) 2026 IEEE. Personal use of this material is permitted. However, permission to use this material for any other purposes must be obtained from the IEEE by sending a request to pubs-permissions@ieee.org.}

}

\begin{document}

\maketitle
\begin{abstract}
  The integration of satellite communication networks with next-generation (NG) technologies is a promising approach towards global connectivity. 
However, the quality of services is highly dependant on the availability of accurate channel state information (CSI). 
Channel estimation in satellite communications is challenging due to the high propagation delay between terrestrial users and satellites, which results in outdated CSI observations on the satellite side. 
In this paper, we study the downlink transmission of multiple satellites acting as distributed base stations (BS) to mobile terrestrial users.
We propose a multi-agent reinforcement learning (MARL) algorithm which aims for maximising the sum-rate of the users, while coping with the outdated CSI. We design a novel bi-level optimisation, procedure themes as dual stage proximal policy optimisation (DS-PPO), for tackling the problem of large continuous action spaces as well as of independent and non-identically distributed (non-IID) environments in MARL. Specifically, the first stage of DS-PPO maximises the sum-rate for an individual satellite and the second stage maximises the sum-rate when all the satellites cooperate to form a distributed multi-antenna BS. Our numerical results demonstrate the robustness of DS-PPO to CSI imperfections as well as the sum-rate improvement attached by the use of DS-PPO. In addition, we provide the convergence analysis for the DS-PPO along with the computational complexity.

\end{abstract}

\begin {IEEEkeywords}
LEO Satellite Communication, Non-terrestrial Networks, Outdated CSI, Multi-Agent Reinforcement Learning, Distributed MIMO
\end{IEEEkeywords}

\section{Introduction}\label{Section:Intro}
\IEEEPARstart{N}{on-terrestrial} networks (NTN) and particularly satellite communications (SatComs) are emerging technologies that are expected to find their way in the next-generation (NG) communication systems. Low-earth orbit (LEO) satellites provide extended coverage for numerous devices, while maintaining a relatively low latency compared to geostationary satellites \cite{10475582}. Examples like the Starlink mega-constellation are already providing a dense network of satellites, enabling users roaming in off-the-grid areas to stay connected \cite{10373866,OmidSpaceMIMO}. The high-altitude platforms, such as unmanned aerial vehicles, are unable to provide continuous, long-term connectivity. Thus, the integration of terrestrial networks with satellites is a vital step for NG systems.

Integrating SatComs with terrestrial networks poses significant challenges in achieving high-throughput connectivity. Some of these challenges include high Doppler shifts, frequent satellite handovers, and long propagation delays \cite{shadab2024}. 
Specifically, propagation delays pose challenges for the channel estimation process in the uplink, as the delay exceeds the channel's coherence interval. To estimate the CSI, users transmit unique pilot sequences, allowing the satellites to estimate the channels between the two parties. However, the propagation delay results in a mismatch between the estimated channel state information (CSI) and the actual CSI. This issue is known as the ``outdated/delayed CSI problem," which we aim to address.

\textcolor{black}{In the literature, several authors have considered the effect of delayed CSI in downlink satellite-user transmission. 
The authors of \cite{9439942} proposed a deep learning (DL)-based satellite channel predictor that used long short-term memory units for eliminating the detrimental impact of the outdated CSI.
In \cite{9826890}, a pair of deep neural networks (DNN) were proposed, one for satellite channel prediction  (SatCP) and another for satellite hybrid beamforming (SatHB). In SatCP, the correlation between the uplink and the downlink CSI was exploited for predicting the downlink CSI by a DNN. By feeding the output of the SatCP network into the SatHB network, the SatHB maps the predicted CSI to a desired hybrid beamformer. 
Furthermore, in \cite{OurPaperTCOM}, an RL algorithm using the deep deterministic policy gradient (DDPG) technique is introduced for coping with the effects of delayed CSI. Unlike previous research, in \cite{OurPaperTCOM}, the delayed CSI is directly mapped to the optimised transmit precoding matrix (TPM), skipping over the channel prediction entirely.}

In LEO satellite networks, the dense constellation of satellites ensures that multiple satellites are visible to terrestrial users at any given time \cite{10646360}. This allows the satellites to operate as a distributed multiple-input multiple-output (MIMO) base station (BS), providing significant diversity gain to improve the throughput, enhance reliability, and mitigate fading effects. These satellites are connected by inter-satellite links, enabling them to share their information effectively \cite{9939157}. To maximise throughput, the satellites must collaboratively form a distributed TPM, while accounting for the challenges of outdated CSI. 
\textcolor{black}{To the best of our knowledge, the consideration of outdated CSI in a cooperative multi-satellite system has only been discussed in [10], where the authors harnessed the statistical information of the channel estimation errors to cope with the deleterious effects of outdated CSI. The simplicity of the technique makes it compatible with systems having low operating frequencies of up to $1$ GHz (see Figure 2 of \cite{10639150})}. 
While it is theoretically possible to model such delays probabilistically and apply robust or stochastic optimisation \cite{6996028}, in practice, high user/satellite mobility and the increasing carrier frequency (e.g., above 1 GHz) introduce rapidly time-varying channels. As shown in \cite{10639150}, the error induced by delayed CSI under these conditions becomes excessive to model accurately via a simple statistical distribution. This renders conventional convex optimisation approaches ineffective due to high statistical uncertainty. For higher operating frequencies, the current literature of multi-satellite systems either considers the unrealistic scenario of perfect CSI or they only rely on statistical CSI.

To fill the knowledge gaps, we aim for directly mapping the delayed CSI to an optimised  TPM in a multi-satellite system. However, the technique conceived in \cite{OurPaperTCOM} cannot be applied here due to the independent and non-identically distributed (non-iid) nature of the environment, where each satellite's path uniquely determines its CSI. Hence, the RL agent(s) must track and adapt to the dynamic changes across multiple channels, significantly increasing the complexity of the task.
In this paper, we explore the potential of multi-agent reinforcement learning (MARL) in counteracting the outdated CSI challenge in cooperative multi-satellite systems, where satellites act as distributed agents. \textcolor{black}{Existing MARL methods that rely on global state sharing or centralised critics may impose substantial inter-satellite communication overhead, since they require the exchange of full CSI, actions, or policy-related information among satellites. This overhead is particularly challenging in LEO multi-satellite systems due to the large continuous action space, propagation delays, and the non-IID nature of the local satellite observations. To address this limitation, DS-PPO reduces the amount of exchanged information by sharing only the singular values of the first-stage TPMs, rather than the full TPMs or global CSI. This compact representation provides coordination information for the second-stage agents, while limiting the inter-satellite signalling overhead.}
Our contributions are as follows:
\begin{enumerate}
    \item We address the challenge of channel aging in multi-satellite downlink communication to mobile devices.  In constrast to \cite{9439942,9826890}, our work bypasses channel prediction altogether by directly mapping the delayed CSI to the TPMs. Furthermore, in contrast to \cite{10639150}, our approach is specifically designed for high-frequency scenarios and incorporates distributed TPM optimisation. This eliminates the reliance on a high-performance network controller and effectively distributes the processing load across satellites.
    \item We propose a novel algorithm termed as Dual-Stage Proximal Policy optimisation (DS-PPO). Unlike the method in \cite{OurPaperTCOM}, the DS-PPO algorithm is specifically designed for cooperative multi-satellite communications with non-iid environments. We introduce a bi-level optimisation framework for handling the complexity of the environment: In a given satellite, the first proximal policy optimisation (PPO) optimises the satellite's TPM to maximise its individual sum-rate. The second PPO leverages the singular values of the first stage TPM across other satellites to  optimise its TPM as part of a distributed satellite system. Sharing these singular values limits the information exchanged among satellites, enabling them to handle the non-iid environment through distributed learning.
    \item \textcolor{\revisioncolor}{We provide the computational complexity and the convergence analysis of the DS-PPO, proving that there is a lower bound on the Stage 2 performance and it is bounded relative to Stage-1, and that DS-PPO can be considered as a light weight algorithm.}
    \item We demonstrate the robustness of DS-PPO to CSI delays and its superior performance in handling complex multi-satellite scenarios through numerical evaluations.
\end{enumerate}

As for the rest of this paper, the system model and problem formulation are presented in Section \ref{sec:ProblemFormulation}. Section \ref{sec:Methodology} explains the proposed algorithm in details. The numerical results are given in Section \ref{sec:results}, and  Section \ref{sec: Conclusions} concludes the paper.

\begin{figure}[!t]
    \centering
    \includegraphics[width=\linewidth]{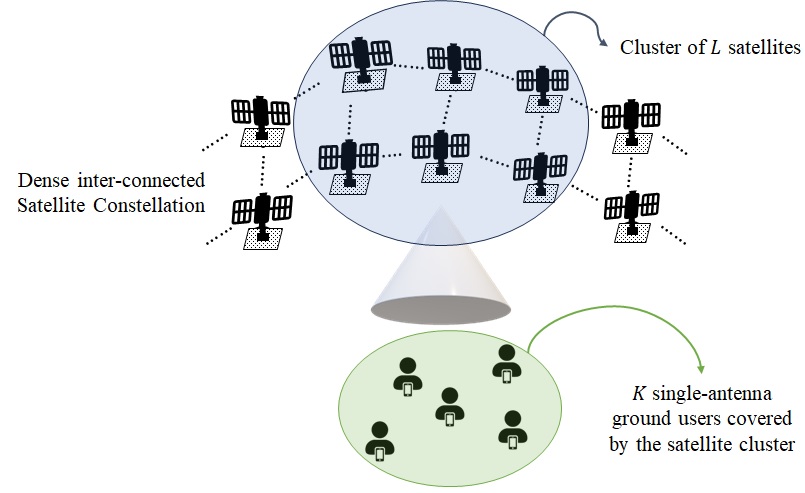}
    \caption{System model: a cluster of $L$ satellites providing service for $K$ users.}
    \label{fig:PPO_with_LLM}
    
\end{figure}

\section{System Model and Problem Formulation} \label{sec:ProblemFormulation}

In this paper, we consider a dense LEO satellite constellation where $L$ satellites are collaborating with each other through their intersatellite links to provide connectivity for $K$ single-antenna mobile users on the ground. Each satellite is equipped with a uniform planar array (UPA) of $M=M_x\times M_y$ antennas. Since in SatComs the channels mostly consist of line of sight (LOS) links, this collaboration among satellites increases the diversity overall, which can enhance the system throughput. 
At each time instant, these satellites aim for transmitting a vector of symbols $\mb{s}(t)=[s^{[1]}(t),...,s^{[K]}(t)]^T$ to the users, and they precode this data via a TPM formulated by $\mb{V}(t,f)=[\mb{v}^{[1]}(t,f),...,\mb{v}^{[K]}(t,f)]\in\mathbb{C}^{ML\times K}$. Note that each $M$ rows of this matrix is in fact a precoding sub-matrix for an individual satellite. In other words, $\mb{V}_l(t,f)\in\mathbb{C}^{M\times K}$ is a sub-TPM for the $l$th satellite, which contains the rows $[(l-1) M+1]$ to $(M l)$ of the matrix $\mb{V}(t,f)$.

\textcolor{black}{A cluster is referred to as the set of satellites that are collaborating with each other for data transmission and form a distributed MIMO system. We use a LEO constellation of satellites similar to Starlink. As in any other LEO constellation, the satellites orbit the earth with a high speed (in the range of 7 km/sec for Starlink). This means that a given satellite has a short window of connection to a given ground user (in the range of 5 minutes). We say a satellite is visible to a ground user when it is within the user's 30 degree elevation angle. At any given time, for a user located in the UK, around 40 satellites are in sight. At a given time $t$, we select $L$ satellites from the set of satellites that are visible to a certain coverage region on earth, to form a cluster. These $L$ satellites are selected based on their distance to the centre of the coverage area, i.e. the $L$ satellites that are closest to the centre of the coverage area at time instant $t$ are selected to form the cluster. However, due to satellites' mobility, these satellites move away from the coverage region and they are replaced in the cluster by satellites with shorter distances to the users. This is called a satellite handover. We use the distance metric for satellite handovers rather than the received power or channel gain, to avoid performing a handover due to short-term channel fluctuations. Instead we base our handover mechanism on the free-space path-loss (FSPL). }

The satellites in a cluster are aiming for optimising their local TPMs in order to maximise the overall sum-rate. 
Thus, we form our problem with the objective of maximising the sum-rate of the users, as 
\begin{equation}\label{main opt}
\begin{array}{cl}
\max\limits_{\mathbf{V}(t,f)} &\displaystyle \sum_{k=1}^{K}\log\Big(1+\frac{|{\mathbf{h}}^{[k]H}(t,f)\mathbf{v}^{[k]}(t,f)|^2}{\sum_{i\neq k}^{K}|{\mathbf{h}}^{[k]H}(t,f)\mathbf{v}^{[i]}(t,f)|^2+\sigma^2} \Big)\\ 
s.t. &\ \text{Tr} \left(\mathbf{V}_l(t,f)\mathbf{V}_l^H(t,f)\right)\leq P_l, \ l\in\{1,...,L\}, 
\end{array}
\end{equation}
where $\mathbf{h}^{[k]}(t,f)\in\mathbb{C}^{ML\times 1}$ is the CSI of user $k$, $\sigma^2$ is the variance of the additive white Gaussian noise at the user side, and $P_l$ is the power budget for the $l$th satellite. In order to solve this optimisation problem, we need accurate CSI at the satellite side. The users transmit pilot sequences, so that the satellites can estimate the channels, but due to the large propagation delay, the CSI that the satellites estimate become outdated. 
\textcolor{black}{The propagation delay between a given user $k$ and a satellite $l$ is $\tau_d(k,l)=\frac{d_{k,l}}{c}$ where $d(k,l)$ is the distance of the two nodes and $c$ is the speed of light. Assuming that the pilots are transmitted every $\Delta T$ seconds, we could say that the CSI is delayed for $T_d(k,l)$ time instances where $T_d(k,l)=\lceil \frac{\tau_d(k,l)}{\Delta T} \rceil$ is the discrete version of the delay. 
Now, at a given time instant $t$ the satellite $l$ has the CSI at $t-T_d(k,l)$. Also, the maximum discrete inter-satellite link delay is defined by $T_{ISL}=\max_{l,j}\lceil \frac{\tau_{isl}(l,j)}{\Delta t} \rceil$ where the inter-satellite delay between satellites $l$ and $j$ is represented by $\tau_{isl}(l,j)$.
By defining the delay of the cluster of satellites as $T_d=\max_{k,l}(T_d(k,l))+T_{ISL}$, we aim to design the TPMs $\mb{V}_l(t,f), \forall l$ by having access to the channels at $t-T_d$, e.g. $\mb{h}^{[k]}(t-T_d,f), \forall k$.} This is referred to as the delayed CSI problem with constant delay. 

\textcolor{black}{In addition, the global sum-rate optimisation problem is non-convex due to the structure of the rate expression. Specifically, the TPM $\mb{v}^{[i]}(t,f)$ (which is an $ML\times 1$ precoding vector from all satellites to the user $i$) appears in both the numerator and the denominator of the SINR expression, and the sum-rate is calculated across users not across the satellites, introducing coupling between agents and dependencies that are hard to isolate. Also, as mentioned above, the CSI available to each satellite is outdated, which can be interpreted as a statistical perturbation of the actual channel (similar to \cite{10639150}). While it is theoretically possible to model such delays probabilistically and apply robust or stochastic optimisation, in practice, high user/satellite mobility and the increasing carrier frequency (e.g., above 1 GHz) introduce rapidly time-varying channels. As shown in \cite{10639150}, the error induced by delayed CSI under these conditions becomes excessive to model accurately via a simple statistical distribution. This renders conventional convex optimisation approaches ineffective due to high statistical uncertainty.}


\section{Background and Preliminaries}
In the following sub-sections we review the basic preliminaries of our proposed algorithm, namely the Augemented Markov Decision Process (MDP) in order to tackle the Delayed CSI, and the Proximal Policy Optimisation (PPO). Then, we discuss the limitations of other novel algorithms in Multi Agent Reinforcement Learning (MARL) in comparison to our problem formulation and enviromental constraints.

\subsection{Augmented Markov Decision Process for Delayed CSI} \label{sub:methodology:cdmdp}
To address the challenge of delayed information, we employ an augmented MDP instead of the traditional MDP. The concept of an augmented MDP was introduced in \cite{altman1992closed}, where the authors propose solutions for managing MDPs with constant delays.
Specifically, the authors suggested that, to construct an equivalent MDP with perfect state information, the agent's previous actions should be included in the state information, as 
\begin{equation}
   \{s({t-T_d}),a({t-T_d}),a({t-T_d+1}),...,a({t-1})\}\in\mathcal{S}^{'},
\end{equation}
where $s({t-T_d})$ is the delayed observation, and $a({t-T_d}), a({t-T_d+1}),..., a({t-1})$ are the previous actions during the delay period.

\subsection{Proximal Policy Optimisation}
PPO is an on-policy, model-free RL algorithm designed to take smaller steps using the most recent available data \cite{SchulmanWDRK17}. PPO addresses two key challenges: determining the appropriate step size for policy updates and maintaining training stability. 
It builds upon concepts used in trust region policy optimisation \cite{SchulmanLMJA15}.
The core idea behind PPO is to make gradual improvements during policy updates \cite{SchulmanWDRK17}. 
It has the ability to optimise the policy using smaller steps and converging faster than other methodologies, thus, it is deemed for our problem formulation with a non-iid distribution and delays.
PPO has a critic network ($V_{\phi}$) to estimate the value function and an actor network ($\pi_{\theta}$) that represents the policy that selects actions.
During each iteration, the algorithm goes through a phase known as "rollout," where the current policy samples actions over $T$ time steps and computes the rewards. At the end of the rollout, PPO calculates the advantage using generalized advantage estimation (GAE), which is given by
\begin{gather}
    \hat{A}^{\pi}_t = \delta_t + (\gamma \lambda)\delta_{t+1} + \cdots + (\gamma\lambda)^{(T-1)}\delta_{T-1},
\end{gather}
where $r_t$ is the reward function, $\gamma$ is the discount factor, $\lambda$ is the GAE discount factor and $\delta_t$ is the advantage function that estimates the advantage of an action $a$ in state $s$ under policy $\pi$ and is given by,
\begin{equation}
A^\pi(s,a) = Q^\pi(s,a) - V^\pi(s)
\end{equation}
where $Q^\pi(s,a) = r(s,a) + \gamma \mathbb{E}_{s' \sim P(\cdot|s,a)}[V^\pi(s')]$ is the action-value function.

Next, the rollout data is divided into small mini-batches, which PPO iterates over for $K$ epochs. During the update epochs, we calculate the ratio between the new policy and the old policy. This ratio is then used to compute the loss for the clipped surrogate objective, which is given by
\begin{equation} \label{eq:rl_clipped_surrogate_function}
    L_{clip} (\theta) = \hat{\mathbb{E}}_t [\mathop{min}(r_t(\theta))\hat{A}^\pi_t, \text{clip}(r_t(\theta), 1 - \epsilon, 1 + \epsilon)\hat{A}^\pi_t]
\end{equation}
where, $\epsilon$ balances the exploration and exploitation of the clipped surrogate function.
Then,  the total loss is computed as the sum of three components: the clipped surrogate objective, the value function loss, and the entropy bonus, which is given by
\begin{equation} \label{eq:rl_ppo_total_loss_function}
    L_{total} = \hat{\mathbb{E}}_t[L_{clip}(\theta) - c_1 \cdot L_{VF}(\theta) + c_2 \cdot \mathbb{H}(\pi_{\theta} \vert s_t) ],
\end{equation}
where, $c_1, c_2$ are the coefficients, and $\mathbb{H}(\cdot)$ is the entropy function and $L_{VF}$ is given by $L_{VF}=(V_\theta(s_t) - V_t^{target}(s_t))^2$, where $V_{\theta}$ and $V_{target}$ are the current value function and the target value, respectively. We use the combined loss in (\ref{eq:rl_ppo_total_loss_function}) in order to update the parameters using the Adam optimiser. With this technique, PPO balances exploration, stability, and learning efficiency. 

\subsection{Limitations of Existing MARL Approaches}
There are several prominent multi-agent reinforcement learning algorithms, which are not directly applicable to our problem formulation due to fundamental architectural and environmental constraints \cite{marl-book, zhang2021multi}. For example, Multi Agent Deep Deterministic Policy Gradient (MADDPG) \cite{lowe2017multi} employs a centralised training with decentralised execution, where each agent's critic requires access to the observations and actions from all agents during training. In our LEO satellite network, this would necessitate gathering $\mathcal{O}(LMK)$ dimensional state and action information from all $L$ satellites at a central location, incurring a large inter-satellite communication overhead. This becomes more problematic considering the large action space in our system. Similarly the Multi Agent Proximal Policy Optimisation (MAPPO) \cite{yu2022surprising} relies on a centralised value function that conditions on global state, which is infeasible when satellites cannot share their full CSI in a timely manner. In QMIX (Monotonic Value Function Factorisation) \cite{rashid2018qmix} and Counterfactual Multi-Agent Policy Gradients (COMA)\cite{foerster2018counterfactual}, it is assumed that there is access to a mixing network or a centralised critic that combines individual agent utilities, requiring synchronous global information exchange that violates the distributed nature of our system. Moreover, these algorithms assume stationary environments where the global policy converges to a fixed equilibrium. For example, recent work on fully decentralised MARL with networked agents \cite{zhang2018fully} has established convergence guarantees for actor-critic algorithms with linear function approximation, but assumes stationary environments which do not apply to our scenario. 

By contrast, the LEO multi-satellite network considered exhibits inherent non-stationarity due to satellite mobility, user mobility, time-varying delayed channels, and frequent handovers that change the serving cluster composition. As a result, the global CSI, interference patterns, and contribution of each satellite to the sum-rate vary continuously rather than being drawn from a fixed distribution. This makes centralised value-function learning challenging, since a centralised critic must approximate a value function over a state-action space whose distribution changes with the cluster geometry, leading to high-variance value estimates, unstable policy-gradient updates, and difficult convergence in practice.\textcolor{\revisioncolor}{Recent works in satellite communications apart from the channel estimation and predicting TPM such as \cite{10720868, 11369878, 11155152} they have used MARL or decentralised algorithms to tackle offloading or resource allocation. For example, in \cite{11155152}, the authors are treating each beam as an agent in a single satellite, and their state and action space comparable with our problem formulation remains relatively small. In addition, in \cite{10720868}, the authors use MARL for offloading and resource allocation for edge computing in satellites, however, the problem formulation is a convex optimisation which is incompatible with our paper. In \cite{11369878}, the authors propose a decentralised framework for radio access technologies, where they reduce the overhead, however, the algorithm is applied in the access control layer rather than in the physical layer.}

\section{Proposed DS-PPO Algorithm}\label{sec:Methodology}
In this section, we discuss the methodology used for solving the problem in (\ref{main opt}).
By exploiting the advantages of PPO, we conceive a bi-level optimisation procedure for first optimising each satellite's individual sum rate and then optimise the overall sum rate for the scenario when the satellites work as a distributed MIMO. By taking into account that the CSI observations are delayed, this problem becomes a constant delay Markov decision process (MDP), which is discussed in the following sub-section. Note that the value of this constant delay in observations is dependent on both the propagation delay and on the pilot transmission rate, which determines the duration of each time step.

\begin{figure}[t]
    \centering
    \includegraphics[width=\linewidth]{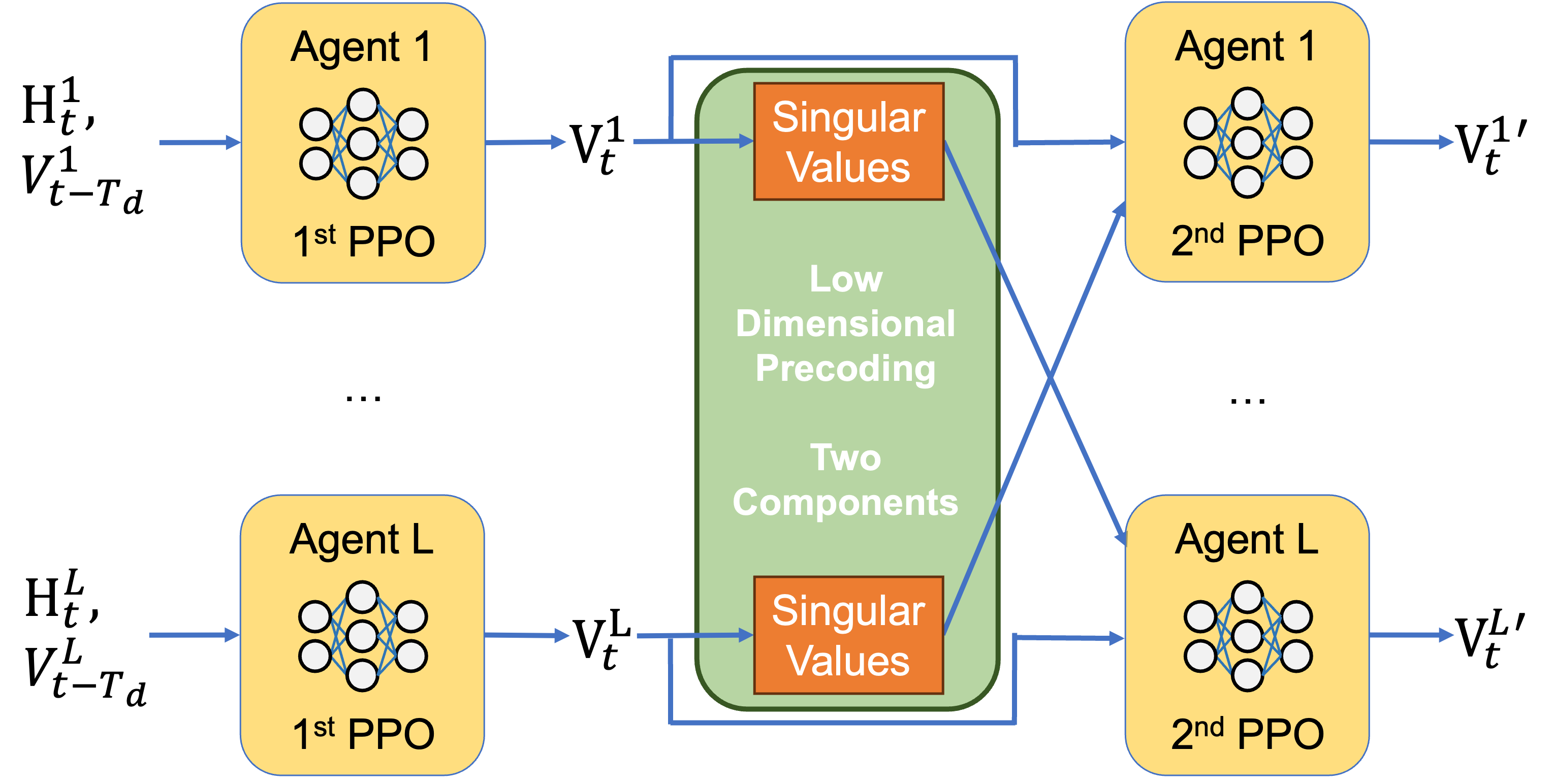}
    \caption{DS-PPO for enhanced TPM generation.}
    \label{fig:DS-PPO_model}
\end{figure}

\subsection{Algorithm}
\textcolor{black}{In our MARL model, each agent represents a satellite $l$ from a cluster of satellites $L$ and it is tasked with calculating the precoding sub-matrix $V^l_t$ specific to its satellite. This computation is performed using information shared by other satellites within the cluster, enabling efficient coordinated operation.}
The DS-PPO MARL algorithm we developed is an evolved version of the PPO RL algorithm defined in \cite{SchulmanWDRK17}. The proposed algorithm operates in two stages to optimise the TPM based on the delayed CSI for a distributed multi-satellite communication system, as shown in Figure 2. This DS-PPO MARL algorithm is particularly useful for distributed learning scenarios where agents are dealing with non-IID environments. 

In the first stage, the algorithm determines the initial TPM for each satellite $V^{l}_t$ such that $\{ V^{1}_t, ... , V^{L}_t\}$, through employing PPO RL  of\cite{SchulmanWDRK17} to optimise each satellite's sum-rate based on its own CSI $H_t^l$ and reward structure. In addition, apart from CSI $H_t^l$ we feed to the agent the action for the particular CSI, i.e, $V_{t-T_d}^l$, \textcolor{black}{with $T_d$ including both user-satellite and inter-satellite delays}. We assume that each satellite is transmitting to the users independently.
Next, the algorithm calculates the singular values of the individual TPMs obtained in the first stage. These singular values are then shared among the satellites in the cluster, providing compact and meaningful information about each satellite’s transmission characteristics. This stage helps reduce the dimensionality of the information exchanged among satellites, making the process more efficient. Additionally, in the first stage of DS-PPO, sharing singular values enables each satellite to assess how much power other satellites are allocating to each user individually. This helps the agents to track the pattern of changes in the CSI of each satellite.

\begin{algorithm}[!h]
\SetKwInOut{Input}{Input}\SetKwInOut{Output}{Output}
\SetKwComment{Comment}{$\triangleright$\ }{}
\Input{CSI $H$}
\Output{Optimal TPMs $V^\prime$}
Let $V_t = \{ V_t^1, ... , V_t^L\}$, $V^\prime_t = \{ V^{\prime 1}_t, ... , V^{\prime L}_t\}$, and $V_{t-T_d} = \{ V_{t-T_d}^1, ... , V_{t-T_d}^L\}$ \\
Let $H_t = \{ H^1_t, ... , H^L_t \}$\\
\For{iteration = 1, 2, ...}{
    \For{t = 1, ..., T}{
        $V_t \leftarrow$ First Stage Agents ($H_t$, $V_{t-T_d}$) \\
        Calculate Reward for $(V_t, H_t)$\\
        Find the singular values of $V_t$ to form $\Lambda$\\
        $V_t^\prime \leftarrow$ Second Stage Agents $(V_t, \Lambda)$\\
        
    }
    Perform PPO Training for both stages\\
}
\caption{DS-PPO Algorithm for MARL}\label{alg:ppo}
\end{algorithm}

\textcolor{black}{In the second stage, the algorithm updates the TPMs $V^{\prime , \ l}_t$ by incorporating the shared singular values and the first stage TPM $V_t^l$, treating the satellites as part of a distributed MIMO system.} An additional agent is trained for each satellite using the second stage of PPO, aiming for maximising the global sum-rate for the entire satellite network. This cooperative update allows the system to track satellite movements and adapt to changes in the CSI, improving the overall transmission performance. The DS-PPO method is given in Algorithm \ref{alg:ppo}

\subsection{State Space}
Due to the fact that there are two different stages in DS-PPO, we define two different state spaces and reward functions. The state space in the first stage uses the augmented state from Section \ref{sub:methodology:cdmdp}. To reduce both the computational complexity and the resources in each satellite, we have selected the pilot transmission rate for ensuring that the CSI observed would have only a single time step transmission delay. In other words, the time duration between two consecutive CSI observations matches the propagation delay of the environment. Thus, the state space of the first stage $s_{t, 1}^l$ at time step $t$ for the $l$-th satellite is 
\begin{equation}\nonumber 
    s^l_{t,1} = \{\text{H}_l(t-T_d,f), \text{V}_l(t-T_d,f)\}
\end{equation}
where $\text{H}_l(t-T_d,f)\in\mathbb{C}^{M\times K}$, is the outdated CSI of the $l$th satellite, while, the $\text{V}_l(t-T_d,f)$ is its TPM. 

Before progressing to the second stage, the satellites share their singular values with the other satellites. In particular we use the squared singular values of their first-stage TPMs with the other satellites in the cluster. \textcolor{\revisioncolor}{For the complex TPM $\mathbf{V}_l(t,f)$, these are obtained as the eigenvalues of the Hermitian positive matrix such as, 
\begin{equation}
    \sigma_{l,i}^2 = \lambda_i\!\left(\mathbf{V}_l(t,f)\,\mathbf{V}_l^H(t,f)\right),
    \quad i = 1,\dots,K,
\end{equation}
where $\lambda_i(\cdot)$ denotes the $i$-th eigenvalue and $\sigma_{l,i}$ is
the $i$-th singular value of $\mathbf{V}_l(t,f)$. We share the squared singular values
$\boldsymbol{\lambda}_l = [\sigma_{l,1}^2,\dots,\sigma_{l,K}^2] \in \mathbb{R}^{1 \times K}$}

Therefore, the state space $s_{t, 2}$ of the second stage is formed by previous stage's action and the singular values from the other satellites in the cluster and it is defined as,
\begin{equation}\nonumber 
s^l_{t, 2} = \{\mathbf{V}_l(t-T_d,f),  \Lambda^l \}, \  \Lambda^l = \bigcup_{j \neq l}^L \lambda_j , \lambda_j\in\mathbb{R}^{1 \times K}
\end{equation}
where $\Lambda^l$ is the set of all singular values \textcolor{\revisioncolor}{of $\mathbf{V}_l(t-T_d,f), \forall l'\neq l$ gleaned from the other satellites in the cluster}.

\subsection{Action Space}
The action space is a continuous space, defined by the real and imaginary components of the TPM coefficients. Thus, its dimensionality is determined by $2KM$.
The action spaces for the first and second stages at time step $t$ for the $l$th satellite are denoted by,
\begin{gather}
    \text{Stage 1:} \quad a^l_{t,1} = \{\mb{V}_l(t,f)\} \nonumber \\
    \text{Stage 2:} \quad a^l_{t,2} = \{\mb{V}_l^\prime(t,f)\} \nonumber 
\end{gather}

\subsection{Reward function}
For each stage we use two different reward function. The first stage reward function is aiming to help the agents to learn and optimise the TPM individually in order to maximise their sum rate and the second stage reward function to incorporate the optimisation objective to optimise the TPM collaboratively.

\subsubsection{First Stage} 
Our reward function consists of three components: $f(\cdot)$, which is associated with the current sum-rate; $g(\cdot)$, which considers both the current and past sum-rates; and a term representing the power constraint. The reward $r^l_t$ for the first stage of PPO, at time step $t$ for the $l$-th satellite is given by
\begin{equation}
    r^l_t = f(c_t) + g(c_t, c_{t-1}) - p_l
\end{equation}
where $c_t$ is the sum rate of users at time $t$ in case only the satellite $l$ is providing service for the users individually. The functions $f(c_t)$, $g(c_t, c_{t-1})$, and $p_l$ are denoted by
\begin{align}
    f(c_t) = & \begin{cases}
        -2, & c_t \leq \xi_1 \\
        -1, & \xi_1 < c_t \leq \xi_2 \\
        0, & \xi_2 < c_t \leq \xi_3 \\
        1, & \xi_3 < c_t \leq \xi_4 \\
        2, & \xi_4 < c_t \leq \xi_5 \\
        2 + \lceil c_t \rceil - \xi_5 , & c_t > \xi_5
    \end{cases} 
\end{align}
\begin{equation}
        g(c_t, c_{t-1}) = \text{sign}(c_t - c_{t-1}) \cdot 1.5 - 0.5,
\end{equation}
\begin{equation}\label{p_l function}
    p_l =\frac{3}{10} \left\lvert tr\left(\mathbf{V}_l(t,f) \cdot \mathbf{V}_l(t,f)^H\right) - P_l \right\rvert,
\end{equation}
with $P_l$ representing the power budget for the $l$th satellite. Note that in order for the agent in satellite $l$ to calculate $c_t$ for the second stage, all satellites in the cluster need to share their CSI observations and their final precoding matrices with each other.

The first part of the reward function $f(c_t)$ is designed in a quantized form to support the learning process of the RL agents. The choice of hyperparameters for this quantization was guided by the target sum-rate we wanted to achieve. In particular, our goal was to ensure a minimum guaranteed sum-rate of around $300$ Mbps, and thus the following parameters in Table~\ref{tab:parameters}. Besides focusing on the sum-rate itself, we also take into account whether the sum-rate has improved during training. For this, we use the function $g(c_t, c_{t-1})$ which simply adds or subtracts a constant from the reward depending on whether the sum-rate has increased compared to the previous time step. Finally, to regulate the power consumption of the TPM, we include the term $p_l$.
\begin{table}[h]
    \centering
    \begin{tabular}{|c|c|c|c|c|}
        \hline
        $\xi_1$ & $\xi_2$ & $\xi_3$ & $\xi_4$ & $\xi_5$  \\
        \hline
        $120$ & $ \frac{5 \xi_1}{4} = 150$ &$ \frac{7\xi_1}{4}= 210$ & $\frac{9\xi_1}{4}= 270$ & $3\xi_1 = 360$ \\
        \hline
    \end{tabular}
    \caption{Parameters used in $f(c_t)$, that ensure a minimum guaranteed sum-rate of around $300$ Mbps }
    \label{tab:parameters}
\end{table}
Note that in (4), with the $P_l$, we represent the power budget for the $l$th satellite to regulate the power consumption of the TPM. The addition of power constraint $p_l$ aims to optimise the TPM such that to maintain the TPM inside the feasibility region. To guarantee this we have project the predicted TPM $\mathbf{V}_l$ into a sphere of radius $P$ \cite{OurPaperTCOM}. \textcolor{black}{The spherical projection in Proposition 1 is applied as a post-processing step on the action by the actor network, and does not interfere with PPO's gradient-based policy updates.}

\begin{proposition}[Power Budget Projection] Given the TPM $\mathbf{V}_l$, the projection of $\mathbf{V}_l$ into a set $\mathbb{S}$  is given  by,
\begin{equation}
\min\limits_{\mb{P} \in \mathbb{S}}\|\mathbf{V}_l-\mb{P}\|
\end{equation}
Given that $\mathbb{S}$ is a sphere centred at the origin with radius $P$, such that ($\mathbb{S}=\{\mathbf{V}_l |\|\mathbf{V}_l\| \leq P\}),$ the projection of $\mathbf{V}_l$ is obtained by
\begin{equation}
P \times \frac{\mathbf{V}_l}{\|\mathbf{V}_l\|+\max (0, P-\|\mathbf{V}_l\|)}    
\end{equation}
\end{proposition}

\textcolor{black}{Also, we note that, the threshold values in Table I are not universal constants. They are design parameters selected according to the propagation environment of the considered Starlink-like LEO constellation. The same thresholds are valid for different values of $\{K,L\}$, in the same system model, however, for a  different constellation or service scenario, the satellite motion, path-loss, Doppler, delay, visibility, and handover patterns will change the distribution of achievable rate. In that case, the thresholds should be recalibrated during the training phase.}

\subsubsection{Second Stage} 
The reward function for the second stage is a simplified representation of our objective, incorporating the logarithm of the sum-rate of users being served by the satellite cluster, denoted as $c^{'}_t$, and the power constraint $p_l$, given by
\begin{align}
    r_t = & 
    \begin{cases}
    log(c^{'}_t) - p_l - 1, \ c^{'}_{t-1} > c^{'}_t, \\
    log(c^{'}_t) - p_l + 1, \ c^{'}_{t-1} < c^{'}_t.
    \end{cases}. 
\end{align}
Note that the logarithmic function is used here to constrain the variations in the reward function, ensuring more stable and manageable updates during optimisation.

\section{\textcolor{black}{Convergence Analysis}}
In this section, we provide the convergence analysis for the DS-PPO. Our convergence analysis builds upon the policy mirror descent framework \cite{lan2023policy}, which establishes linear convergence rates for policy optimisation methods. Given the fact that capacity depends on the state of Stage 1, and action of State 2, we have considered how the action of Stage 2 converged to the optimal solution and improves the Stage 1.

To recap, we model each satellite's decision process as a Markov Decision Process (MDP) $\mathcal{M} = (\mathrm{S}, \mathrm{A}, P, r, \gamma, \rho_0)$ where $\mathrm{S}$ is the state space, $\mathrm{A}$ is the action space, $P: \mathrm{S} \times \mathrm{A} \to \Delta(\mathrm{S})$ is the transition kernel, $r: \mathrm{S} \times \mathrm{A} \to \mathbb{R}$ is the reward function, $\gamma \in [0,1)$ is the discount factor, and $\rho_0 \in \Delta(\mathrm{S})$ is the initial state distribution \cite{sutton2018reinforcement}. Also, note that $\pi_1$ and $\pi_2$ are the policies for the first and second stage, respectively.

Our analysis follows the framework established in \cite{agarwal2021theory} for analysing policy gradient convergence. The following definitions are important for the convergence analysis and prove that stage 2 improves the policy of stage 1 and reaches an optimal solution.

\begin{definition}[Policy Performance]
The expected cumulative discounted reward under policy $\pi$ is:
\begin{equation}
\eta(\pi) = \mathbb{E}_{\tau \sim \pi}\left[\sum_{t=0}^{\infty} \gamma^t r(s_t, a_t)\right] = \mathbb{E}_{s_0 \sim \rho_0}[V^\pi(s_0)]
\label{eq:eta_def}
\end{equation}
where $V^\pi(s) = \mathbb{E}_\pi[\sum_{t=0}^\infty \gamma^t r(s_t, a_t) | s_0 = s]$ is the value function.
\end{definition}

\begin{definition}[Advantage Function]
The advantage of action $a$ in state $s$ under policy $\pi$ is:
\begin{equation}
A^\pi(s,a) = Q^\pi(s,a) - V^\pi(s)
\label{eq:advantage_def}
\end{equation}
where $Q^\pi(s,a) = r(s,a) + \gamma \mathbb{E}_{s' \sim P(\cdot|s,a)}[V^\pi(s')]$ is the action-value function.
\end{definition}

We make the following assumptions in order establish several technical lemmas that will be used in order to provide the upper bound the policy improvement.

\begin{assumption}[Bounded Rewards]
\label{ass:bounded_reward}
There exists $R_{\max} > 0$ such that $|r(s,a)| \leq R_{\max}$ for all $(s,a) \in \mathcal{S} \times \mathrm{A}$.
\end{assumption}
\textcolor{black}{Rewards are bounded the modelling the physical constraint imposed by the power budget and Shannon's models on sum-rate in second stage.}


Based on the above assumptions, we have expressed the following.

\begin{theorem}[Stage~2 Performance Lower Bound]
\label{thm:improvement_main}
Let $\pi_{1}$ be the converged Stage~1 policy and $\pi_{2}$ the Stage~2 policy obtained by DS-PPO, under Assumption 1 \& 2:
\begin{equation}
\eta(\pi_{2}) \geq \eta(\pi_{1}) - \frac{4\epsilon_{\max}\gamma}{(1-\gamma)^2} \cdot D_{TV}^{\max}(\pi_{1}, \pi_{2})^2
\label{eq:improvement_main}
\end{equation}
where $\epsilon_{\max} = \max_{s,a}|A^{\pi_{1}}(s,a)| \leq \frac{2R_{\max}}{1-\gamma}$ is the maximum advantage, and $D_{TV}^{\max}(\pi, \pi') = \max_{s} D_{TV}(\pi(\cdot|s) \| \pi'(\cdot|s))$ is the maximum total variation divergence, under Assumptions 1
\end{theorem}
\textcolor{\revisioncolor}{Equation~\eqref{eq:improvement_main} bounds the performance of the
Stage 2 policy with the Stage 1 that if there is any degradation can be a quadratic in the policy divergence. The  above theorem is consistent with recent results on policy gradient methods \cite{lin2022convergence} and \cite{SchulmanLMJA15}.}

\subsection{Preliminaries}
Before expressing the upper bound for convergence analysis \cite{konda2004convergence}, we have established the following lemmas.
\begin{lemma}[Bounded Value Function \cite{konda2004convergence, lin2022convergence}]
\label{lemma:bounded_value}
Under Assumption~\ref{ass:bounded_reward}, for any policy $\pi$:
\begin{gather}
|V^\pi(s)| \leq \frac{R_{\max}}{1-\gamma}, \quad |Q^\pi(s,a)| \leq \frac{R_{\max}}{1-\gamma}, \\ 
|A^\pi(s,a)| \leq \frac{2R_{\max}}{1-\gamma} \nonumber
\end{gather}
for all $s \in \mathrm{S}$ and $a \in \mathrm{A}$.
\end{lemma}

\begin{lemma}[Performance Difference Lemma \cite{kakade2002approximately}]
\label{lemma:perf_diff}
For any two policies $\pi$ and $\tilde{\pi}$:
\begin{equation}
\eta(\tilde{\pi}) = \eta(\pi) + \mathbb{E}_{\tau \sim \tilde{\pi}}\left[\sum_{t=0}^{\infty} \gamma^t A^\pi(s_t, a_t)\right]
\label{eq:perf_diff_lemma}
\end{equation}
where the expectation is over trajectories $\tau = (s_0, a_0, s_1, a_1, \ldots)$ with $s_0 \sim \rho_0$, $a_t \sim \tilde{\pi}(\cdot|s_t)$, and $s_{t+1} \sim P(\cdot|s_t, a_t)$.
\end{lemma}

\begin{lemma}[Discounted State Visitation]
\label{lemma:visitation}
Define the discounted state visitation frequency under policy $\pi$ as:
\begin{equation}
\rho_\pi(s) = (1-\gamma) \sum_{t=0}^\infty \gamma^t P(s_t = s | \pi, s_0 \sim \rho_0)
\end{equation}
Then $\rho_\pi$ is a valid probability distribution over $\mathrm{S}$, and the performance difference can be written as:
\begin{equation}
\eta(\tilde{\pi}) - \eta(\pi) = \frac{1}{1-\gamma} \sum_{s} \rho_{\tilde{\pi}}(s) \sum_a \tilde{\pi}(a|s) A^\pi(s,a)
\label{eq:perf_diff_visitation}
\end{equation}
\end{lemma}

\begin{lemma}[Coupling Lemma \cite{levin2009markov}]
\label{lemma:coupling}
For any two probability distributions $p$ and $q$ over a finite set $\mathcal{X}$ with total variation distance $D_{TV}(p \| q) = \frac{1}{2}\sum_x |p(x) - q(x)| = \alpha$, there exists a joint distribution $(X, Y)$ with marginals $X \sim p$ and $Y \sim q$ such that:
\begin{equation}
P(X \neq Y) = \alpha = D_{TV}(p \| q)
\end{equation}
\end{lemma}

\subsection{Proof of Theorem~\ref{thm:improvement_main}}
\label{sec:proof_improvement}
The proof follows the approach of Schulman et al. \cite{SchulmanLMJA15} with detailed exposition. We first begin to define the surrogate objective from the PPO. From Lemma~\ref{lemma:perf_diff}, the performance of $\pi_{2}$ relative to $\pi_{1}$ is:
\begin{equation}
\eta(\pi_{2}) = \eta(\pi_{1}) + \mathbb{E}_{\tau \sim \pi_{2}}\left[\sum_{t=0}^{\infty} \gamma^t A^{\pi_{1}}(s_t, a_t)\right] 
\label{eq:exact_perf}
\end{equation}

The difficulty is that the expectation is over trajectories from $\pi_{2}$, which we are trying to optimise. We instead define a \textit{surrogate objective} that uses trajectories from $\pi_{1}$:
\begin{align}
L_{\pi_{1}}(\pi_{2}) & = \eta(\pi_{1}) \\
& + \mathbb{E}_{\tau \sim \pi_{1}}\left[\sum_{t=0}^{\infty} \gamma^t \frac{\pi_{2}(a_t|s_t)}{\pi_{1}(a_t|s_t)} A^{\pi_{1}}(s_t, a_t)\right] \nonumber
\label{eq:surrogate_def}
\end{align}

Using Lemma~\ref{lemma:visitation}, this can be written as:
\begin{align}
L_{\pi_{1}}(\pi_{2}) = & \ \eta(\pi_{1}) \\
& + \frac{1}{1-\gamma} \sum_s \rho_{\pi_{1}}(s) \sum_a \pi_{2}(a|s) A^{\pi_{1}}(s,a) \nonumber
\label{eq:surrogate_visitation}
\end{align}

Here the key observation is when $\pi_{2} = \pi_{1}$, meaning that there is no improvement over the baseline policy (Stage 1), by definition, advantage becomes zero, because the expected advantage of a policy over itself is always zero. Therefore,
\begin{align}
L_{\pi_{1}}(\pi_{1}) = \eta(\pi_{1})
\end{align}

Also, at this point, the gradient will match as below,
\begin{equation}
\nabla_{\pi_{2}} L_{\pi_{1}}(\pi_{2})\Big|_{\pi_{2} = \pi_{1}} = \nabla_{\pi_{2}} \eta(\pi_{2})\Big|_{\pi_{2} = \pi_{1}}
\end{equation}

This means the surrogate $L_{\pi_{1}}$ is a first-order approximation to $\eta$ around $\pi_{1}$.

The next step in the proof is to bound the gap between $\eta(\pi_{2})$ and $L_{\pi_{1}}(\pi_{2})$. The gap arises because the $\eta(\pi_{2})$ uses state visitation $\rho_{\pi_{2}}$ and the $L_{\pi_{1}}(\pi_{2})$ uses state visitation $\rho_{\pi_{1}}$. We first need to define the expected advantage under $\pi_{2}$ at state $s$:
\begin{align}
\bar{A}^{\pi_{1}}(s) & = \mathbb{E}_{a \sim \pi_{2}(\cdot|s)}[A^{\pi_{1}}(s,a)] \\
&= \sum_a \pi_{2}(a|s) A^{\pi_{1}}(s,a) \nonumber
\end{align}

From Lemma~\ref{lemma:visitation} we can express the difference between two policies such as:
\begin{align}
\eta(\pi_{2}) - L_{\pi_{1}}(\pi_{2}) & = \\
& \frac{1}{1-\gamma} \sum_s (\rho_{\pi_{2}}(s) - \rho_{\pi_{1}}(s)) \bar{A}^{\pi_{1}}(s) \nonumber
\end{align}

We now bound $|\rho_{\pi_{2}}(s) - \rho_{\pi_{1}}(s)|$ using coupling arguments. First, let 
\begin{align}
\alpha = D_{TV}^{\max}(\pi_{1}, \pi_{2}) = \max_s D_{TV}(\pi_{1}(\cdot|s) \| \pi_{2}
\end{align}

By Lemma~\ref{lemma:coupling}, at each state $s$, we can couple the actions under $\pi_{1}$ and $\pi_{2}$ such that they differ with probability at most $\alpha$. We consider that there are two trajectories:
\begin{itemize}
    \item $\tau^{(1)} = (s_0, a_{0, 1}, s_{1, 1}, a_{1, 1}, \ldots)$ with $a_{t, 1} \sim \pi_{1}(\cdot|s_{t, 1})$
    \item $\tau^{(2)} = (s_0, a_{0, 2}, s_{1, 2}, a_{1, 2}, \ldots)$ with $a_{t, 2} \sim \pi_{2}(\cdot|s_{t, 2})$
\end{itemize}
starting from the same initial state $s_0$. Using the coupling, we can ensure that at each step, given the trajectories have not diverged:
\begin{equation}
P(a_{t, 1} \neq a_{t, 2} | s_{t, 1} = s_{t, 2}) \leq \alpha \nonumber
\end{equation}

Let $T$ be the first time the trajectories diverge (i.e., $a_{t, 1} \neq a_{t, 2}$ or equivalently $s_{T+1}^{(1)} \neq s_{T+1}^{(2)}$). Then:
\begin{equation}
P(T > t) = P(\text{trajectories identical up to time } t) \geq (1-\alpha)^t \nonumber
\end{equation}

Therefore, the probability that the trajectories have diverged by time $t$ is:
\begin{equation}
P(T \leq t) \leq 1 - (1-\alpha)^t
\end{equation}

Using the inequality $1 - (1-\alpha)^t \leq t\alpha$ for $\alpha \in [0,1]$:
\begin{equation}
P(s_{t, 1} \neq s_{t, 2}) \leq t\alpha
\end{equation}

Now we need to find a relationship in order to bound correctly the state visitation difference. For any function $f: \mathrm{S} \to \mathbb{R}$ with $|f(s)| \leq f_{\max}$:
\begin{align}
&\left|\mathbb{E}_{s_t \sim \pi_{2}}[f(s_t)] - \mathbb{E}_{s_t \sim \pi_{1}}[f(s_t)]\right| \nonumber \\
&= \left|\mathbb{E}[f(s_{t, 2}) - f(s_{t, 1})]\right| \nonumber \\
&= \left|\mathbb{E}[f(s_{t, 2}) - f(s_{t, 1}) | s_{t, 1} \neq s_{t, 2}] \cdot P(s_{t, 1} \neq s_{t, 2})\right| \nonumber \\
&\quad + \left|\mathbb{E}[f(s_{t, 2}) - f(s_{t, 1}) | s_{t, 1} = s_{t, 2}] \cdot P(s_{t, 1} = s_{t, 2})\right| \nonumber \\
&\leq 2f_{\max} \cdot P(s_{t, 1} \neq s_{t, 2}) + 0 \\
&\leq 2f_{\max} \cdot t\alpha
\end{align}

Applying this with $f(s) = \bar{A}^{\pi_{1}}(s)$ and $f_{\max} = \epsilon_{\max} = \max_{s,a}|A^{\pi_{1}}(s,a)|$:
\begin{equation} \label{eq:relAdvantage}
\left|\mathbb{E}_{s_t \sim \pi_{2}}[\bar{A}^{\pi_{1}}(s_t)] - \mathbb{E}_{s_t \sim \pi_{1}}[\bar{A}^{\pi_{1}}(s_t)]\right| \leq 2\epsilon_{\max} \cdot t\alpha
\end{equation}

If we go through the summation over time we can find a bound.
\begin{align}
&\left|\eta(\pi_{2}) - L_{\pi_{1}}(\pi_{2})\right| \nonumber  \\
&= \left|\sum_{t=0}^\infty \gamma^t \left(\mathbb{E}_{s_t \sim \pi_{2}}[\bar{A}^{\pi_{1}}(s_t)] - \mathbb{E}_{s_t \sim \pi_{1}}[\bar{A}^{\pi_{1}}(s_t)]\right)\right| \nonumber  \\
&\leq \sum_{t=0}^\infty \gamma^t \cdot 2\epsilon_{\max} \cdot t\alpha \nonumber  \\
&= 2\epsilon_{\max} \alpha \sum_{t=0}^\infty t \gamma^t
\end{align}

We compute the sum $\sum_{t=0}^\infty t \gamma^t$. Let $S = \sum_{t=0}^\infty t \gamma^t$. Then:
\begin{align}
S &= 0 + \gamma + 2\gamma^2 + 3\gamma^3 + \cdots \nonumber \\ 
\gamma S &= \gamma^2 + 2\gamma^3 + 3\gamma^4 + \cdots \nonumber \\
S - \gamma S &= \gamma + \gamma^2 + \gamma^3 + \cdots = \frac{\gamma}{1-\gamma} \nonumber \\
S &= \frac{\gamma}{(1-\gamma)^2} \nonumber 
\end{align}

Therefore if substitute the above into \ref{eq:relAdvantage}
\begin{align}
\left|\eta(\pi_{2}) - L_{\pi_{1}}(\pi_{2})\right| & \leq 2\epsilon_{\max} \alpha \cdot \frac{\gamma}{(1-\gamma)^2} \nonumber \\
& = \frac{2\epsilon_{\max} \gamma \alpha}{(1-\gamma)^2}
\label{eq:gap_linear}
\end{align}

The bound \eqref{eq:gap_linear} is linear in $\alpha$, and while mathematically valid, a linear bound is insufficient for guaranteeing policy improvement, thus we need refine $\alpha$ for a quadratic. Following \cite{SchulmanLMJA15}, we use the fact that for small $\alpha$:
\begin{equation}
1 - (1-\alpha)^t \leq \min(1, t\alpha)
\end{equation}

For the coupled trajectories, the probability of divergence by time $t$ satisfies:
\begin{equation}
P(T \leq t) = 1 - (1-\alpha)^{t+1} \leq (t+1)\alpha
\end{equation}

A more careful analysis in \cite{SchulmanLMJA15} shows that we can obtain a quadratic form of $\alpha$ thus:
\begin{equation}
\left|\eta(\pi_{2}) - L_{\pi_{1}}(\pi_{2})\right| \leq \frac{4\epsilon_{\max} \gamma}{(1-\gamma)^2} \alpha^2
\label{eq:gap_quadratic}
\end{equation}

The PPO's update maximises the surrogate objective $L_{\pi_{1}}(\pi_{2})$ over policies $\pi_{2}$. Since the Stage 2 update seeks to improve the surrogate:
\begin{equation}
L_{\pi_{1}}(\pi_{2}) \geq L_{\pi_{1}}(\pi_{1}) = \eta(\pi_{1})
\end{equation}

Combining with \eqref{eq:gap_quadratic}:
\begin{align}
\eta(\pi_{2}) &\geq L_{\pi_{1}}(\pi_{2}) - \frac{4\epsilon_{\max} \gamma}{(1-\gamma)^2} \alpha^2 \\
&\geq \eta(\pi_{1}) - \frac{4\epsilon_{\max} \gamma}{(1-\gamma)^2} \alpha^2
\end{align}

With substituting $\alpha = D_{TV}^{\max}(\pi_{1}, \pi_{2})$ completes the proof.

\subsection{Remarks}
The analysis above involves $A^{\pi_{1}}(s,a)$ rather than $A^{\pi_{2}}(s,a)$. This follows directly from the Performance Difference Lemma, which states that for any two policies $\pi$ and $\tilde{\pi}$:
\begin{equation}
\eta(\tilde{\pi}) = \eta(\pi) + \mathbb{E}_{\tau \sim \tilde{\pi}}\left[\sum_{t=0}^{\infty} \gamma^t A^{\pi}(s_t, a_t)\right].
\end{equation}
Thus, the advantage function is always evaluated with respect to the baseline policy $\pi$, not the new policy $\tilde{\pi}$. This is because $A^{\pi}(s,a) = Q^{\pi}(s,a) - V^{\pi}(s)$ measures how much better action $a$ is compared to the average performance under $\pi$.  To evaluate the improvement of $\pi_{2}$ over $\pi_{1}$, we must use $\pi_{2}$'s actions against $\pi_{1}$'s value function which quantifies the improvement relative to $\pi_{1}$. Using $A^{\pi_{2}}$ will provide no information about the improvement of $\pi_{2}$ over $\pi_{1}$. Thus, $\bar{A}^{\pi_{1}}(s)$ captures the expected benefit of following $\pi_{2}$'s action selection while measuring performance gains relative to the Stage 1 baseline.

\textcolor{\revisioncolor}{Also, note that the above analysis is to prove that there is a lower bound so the result guarantees that any performance loss incurred by the Stage 2 update is bounded, and that this bound tightens quadratically as the two policies become close. Additional strict penalty is required in order to prove that there is an improvement.}

\section{\textcolor{black}{Computational Complexity and Memory}}
This section analyses the floating-point operations (FLOPS) and memory required by DS-PPO. We follow standard conventions for counting operations in neural networks and matrix computations \cite{goodfellow2016deep, golub2013matrix}.

\subsection{Complexity}

For matrix operations, we use established results from numerical linear algebra \cite{golub2013matrix}. The matrix multiplication of $C = AB$ where $A \in \mathbb{R}^{m \times n}$ and $B \in \mathbb{R}^{n \times p}$ requires $2mnp$ FLOPS. Each of the $mp$ output elements requires $n$ multiplications and $n-1$ additions). For a neural network layer with input dimension $d_{\text{in}}$, output dimension $d_{\text{out}}$, and batch size $B$, the forward pass requires $2B \cdot d_{\text{in}} \cdot d_{\text{out}}$ FLOPS, while the backward pass for computing the gradients with respect to both weights and inputs, requires double the amount of the forward pass computation\cite{goodfellow2016deep}. Thus, for an MLP with $\ell$ hidden layers, input dimension $d_{\text{in}}$, and output dimension $d_{\text{out}}$, the total forward pass FLOPS are:
\begin{equation}
F_{\text{forward}}^{\text{MLP}} = 2\left(d_{\text{in}} \cdot d_{h_1} + \sum_{i=1}^{\ell-1} d_{h_i} \cdot d_{h_{i+1}} + d_{h_\ell} \cdot d_{\text{out}}\right)
\end{equation}
The backward pass requires approximately $2\times$ the forward pass FLOPS \cite{goodfellow2016deep}, giving a total of $F_{\text{train}}^{\text{MLP}} \approx 3 \cdot F_{\text{forward}}^{\text{MLP}}$ for one training iteration. In our implementation, the actor network uses three layers with 64 neurons each, while the critic network uses layers of 128, 64, and 64 neurons.

For DS-PPO, the input/output dimensions are determined by system parameters:
\begin{align}
\text{Stage 1:} \quad d_{\text{in}}^{(1)} &= 4MK, \quad d_{\text{out}}^{(1)} = 2MK \\
\text{Stage 2:} \quad d_{\text{in}}^{(2)} &= 2MK + (L-1)K, \quad d_{\text{out}}^{(2)} = 2MK
\end{align}
where the Stage~1 input is the complex CSI (real and imaginary parts) and the past action, and Stage~2 additionally receives $(L-1)K$ singular values $\Lambda^l$ from other satellites. The SVD of a complex $M \times K$ matrix requires approximately $F_{\text{SVD}} \approx 8MK^2$ FLOPS \cite{golub2013matrix}. The complexity for each time and single satellite $F_{\text{step}}$ is given by,
\begin{align}
F_{\text{step}} &= F_{\text{actor}}^{(1)} + F_{\text{critic}}^{(1)} + F_{\text{SVD}} + F_{\text{actor}}^{(2)} + F_{\text{critic}}^{(2)} \nonumber \\
F^{(1)} &= F_{\text{actor}}^{(1)} + F_{\text{critic}}^{(1)} \nonumber  \\
F^{(2)} &= F_{\text{actor}}^{(2)} + F_{\text{critic}}^{(2)} \nonumber 
\end{align}
where, $F_{\text{actor}}^{(1)}$ is the Stage 1 actor network,  $F_{\text{critic}}^{(1)}$ is the Stage 2 critic network,  $F_{\text{SVD}}$ is singular values decomposition, $F_{\text{actor}}^{(2)}$ is the Stage 2 actor network, $F_{\text{critic}}^{(2)}$ is the Stage 1 critic network, $F^{(1)}$ is the Stage 1 networks and $F^{(2)}$ is the Stage 2 networks. For the total episode that combines rollout and training is given by,
\begin{equation}
F_{\text{epis}} = \underbrace{T \cdot L \cdot F_{\text{step}}}_{\text{rollout}} + \underbrace{3 \cdot T \cdot L \cdot \left(E^{(1)} \cdot F^{(1)} + E^{(2)} \cdot F^{(2)}\right)}_{\text{training (fwd + bwd)}}
\end{equation}
where $T$ is the total amount of time steps, and $E^{(1)}$, $E^{(2)}$ are the local epochs for stage 1 and stage 2 respectively.

Combining all components, the total training complexity scales as:
\begin{align}
F_{\text{total}} = \mathcal{O}(N_{\text{ep}} \cdot T \cdot L \cdot ( d_h \cdot MK & + d_h \cdot LK \nonumber \\
& + \ell \cdot d_h^2 + MK^2 ))
\end{align}
where $N_{\text{ep}}$ is the number of episodes. For fixed network architecture ($d_h$, $\ell$ constant), this simplifies to:
\begin{align}
F_{\text{total}} &= \mathcal{O}\left(N_{\text{ep}} \cdot T \cdot L \cdot K \cdot (M + L + K)\right)
\end{align}
The dominant terms are: linear in $L$ (satellites), linear in $M$ (antennas), and quadratic in $K$ (users) due to SVD. For small $K$, the neural network terms $\mathcal{O}(MK + LK)$ dominate; for large $K$, the SVD term $\mathcal{O}(MK^2)$ becomes significant. In Table~\ref{tab:flops}, we summarise the FLOPS for different configurations, showing how complexity scales with $L$ and $K$. The dominant cost is neural network training (forward + backward passes), accounting for over 99\% of total FLOPS. The SVD computation contributes less than 1\%.  

\begin{table}[!ht]
\centering
\caption{FLOPS Analysis for DS-PPO ($M = 9$, $T = 512$, 389 episodes, 200K samples)}
\label{tab:flops}
\begin{tabular}{|c|c|c|c|}
\hline
$\bm{L}$ & $\bm{K}$ & \multicolumn{1}{m{2cm}|}{\centering\textbf{Per-Episode (GFLOPS)}} & \multicolumn{1}{m{2cm}|}{\centering\textbf{Total Training (TFLOPS)}} \\
\hline
4 & 2 & 1.7 & 0.66 \\
\hline
4 & 4 & 2.9 & 1.13 \\
\hline
4 & 6 & 4.3 & 1.67  \\
\hline
6 & 2 & 2.6 & 1.01 \\
\hline
6 & 4 & 4.5 & 1.75  \\
\hline
6 & 6 & 6.7 & 2.61  \\
\hline
20 & 30 & 368 & 143.2\\
\hline
\end{tabular}
\end{table}

\subsection{Memory}
\textcolor{black}{Each satellite has four neural networks, an actor and a critic for each of the two stages. The parameter count of a fully connected layer with input dimension of $d_{\mathrm{in}}$ and output dimension $d_{\mathrm{out}}$ is $d_{\mathrm{in}} \cdot d_{\mathrm{out}} + d_{\mathrm{out}}$ (weights plus biases) \cite{goodfellow2016deep}. For a configuration  of $L = 4$, $K = 4$, $M = 9$, the actor at Stage 1 has an input dimension $d_{\mathrm{in}}^{(1)} = 4MK = 144$, an output dimension $d_{\mathrm{out}}^{(1)} = 2MK = 72$, and three hidden layers of 64 neurons each. This yields a total of 22,280 parameters. The critic at Stage 1 has hidden layers of 128, 64, and 64 neurons and a scalar output which yields 31,041 parameters. The actor and critic at Stage 2 have input dimension $d_{\mathrm{in}}^{(2)} = 2MK + (L-1)K = 84$ and contributes 18,440 and 23,361 parameters, respectively. Thus, the total number of parameters per satellite is $N_{\mathrm{params}} \approx 95,100$.}

\textcolor{black}{Considering the use of FP32 precision meaning that we use 4 bytes per parameter, the model weights require $ N_{\mathrm{params}} \approx 372 \text{KB}$\, memory per satellite. The peak activation memory of the largest intermediate tensor during a single forward pass is determined by the widest hidden layer (128 neurons), contributing less than 1 KB. \textcolor{\revisioncolor}{At the inference time only the current observation and the resulting action need to be in the memory. Thus, if we include the input/output buffers (state observations and action outputs for 512 steps, ${\sim}1.5$\,KB)} and SVD storage for the singular value vectors (${\sim}0.3$\,KB), the total inference memory is
\begin{equation}
    \mathcal{M}_{\mathrm{infer}} \;\approx\; 374\;\text{KB per satellite (FP32)}.
\end{equation}}

\textcolor{black}{During the training of the agent additional memory is required. The Adam optimiser requires $2 \times 4 \times N_{\mathrm{params}} \approx 743$ KB of memory at FP32. In addition, each stage maintains its own rollout buffer of $T = 512$ time steps, storing states, actions, rewards, log-probabilities, value estimates, advantages, and returns. For Stage 1, each step stores $d_{\mathrm{in}}^{(1)} + d_{\mathrm{out}}^{(1)} + 5 = 221$ floats which yields $512 \times 221 \times 4 \approx 442$KB memory. For Stage 2, each step stores $d_{\mathrm{in}}^{(2)} + d_{\mathrm{out}}^{(2)} + 5 = 161$ floats, which yields $512 \times 161 \times 4 \approx 322$ KB. Thus, the combined rollout memory is $\approx 764$ KB. We use Pytorch for the backend framework for the algorithm, which allocates gradient tensors of the same size as the model parameters, requiring an additional ${\sim}372$ KB. Therefore, The total training memory per satellite is
\begin{equation}
    \mathcal{M}_{\mathrm{train}} \approx 2.2\;\text{MB per satellite with FP32}.
\end{equation}}

\section{Numerical Results} \label{sec:results}

In this section, we present our numerical results to evaluate the performance of the proposed DS-PPO MARL algorithm. We incorporate a four-layer constellation of $4236$ LEO satellites modelled after Starlink, each equipped with a UPA of $M=9$ antennas. The constellation details are provided in \cite{OurPaperTCOM}. In this constellation, at each instant, several satellites are visible to any user on the ground. The center of the coverage area on Earth has the latitude of $54.5260000^{\circ}$ and the longitude of $-3.3000000^{\circ}$, and the radius of the area is $50$ km.  We select the $L\in\{4,6,8\}$ nearest satellites to the centre of the coverage area to form a cluster. 
As time passes and the satellites move in their orbit, some of the satellites in the cluster may distance themselves from the users and give their position in the cluster to another satellite. The transmission frequency is set to $f=2$ GHz, the bandwidth is set to $BW=0.02f$, and the channel model is the same as in \cite{OurPaperTCOM}. \textcolor{black}{The propagation delay is in the range of $2$ msec based on the specifications of the constellation and the satellite handovers. Also, the ISL delays associated with sharing the singular values must be considered. In our model, the ISL delay in a cluster of $6$ satellites averages around $1$ ms. Given that the pilots are transmitted every $\Delta t=1$ ms, the environment experiences a delay of $T_d=3$ time steps.}

\textcolor{black}{The numerical experiments include both stochastic and time-varying components. A number of $K\in\{2,4,6\}$ single-antenna users are randomly located within the coverage area and move in arbitrary directions at random speeds of up to $3$ m/s. Consequently, the satellite-user distances, Doppler shifts, propagation delays, channel coefficients, and delayed CSI observations vary over time. The channel model also includes stochastic NLOS components, including the number of NLOS paths, complex path gains, NLOS delays, and angular parameters. In addition, the LEO satellite motion changes the serving geometry and leads to handovers and cluster reconfiguration during the simulation. Therefore, the reported episodic sum-rate curves correspond to the learning performance under stochastic user placement, user mobility, time-varying satellite geometry, delayed CSI, and dynamic satellite visibility. For more details on the channel model and satellite handovers please refer to \cite{OurPaperTCOM}. Our results were obtained using a single random seed for each configuration. Nevertheless, the convergence behaviour observed across different values of \(K\), \(L\), and \(T_d\) suggests that the learned policy is robust to the stochastic user mobility, NLOS channel randomness, delayed CSI, and dynamic handovers considered in our simulation setup.}

Both stages of PPO  are run with identical parameters for all agents. The actor network uses the default neural network architecture for PPO, and a Multi-Layer Perceptron (MLP) consisting of three layers with 64 neurons each. The critic network has been modified to include three layers of 128, 64, and 64 neurons, respectively. 
Both stages share this network structure and they use the ADAM optimiser. The hyperparameters for the first and second stages of PPO are listed in Table \ref{tab:ppo parameters}, and they closely follow the guidelines of \cite{shengyi2022the37implementation} with minimal deviations.

\begin{table}[!h]
    \centering
    \caption{PPO Hyper Parameters}
    \begin{tabular}{|c|c|c|c|}
        \hline
        \textbf{Parameters} & \textbf{Symbol} & \textbf{First Stage} & \textbf{Second stage} \\
        \hline
        Discount Factor & $\gamma$ & 0.99 & 0.95 \\
        \hline
        GAE & $\lambda$ & 0.95 & 0.9 \\
        \hline
        Learning Rate & $\eta$ & 0.003 & 0.004 \\
        \hline
        Minibatches & $\gamma$ & 32 & 6 \\
        \hline
        Clip coefficient & $\epsilon$ & 0.1 & 0.01 \\
        \hline
        Maximum Gradient Norm & $\lVert \nabla_{s,a} \rVert$ & 1 & 0.6 \\
        \hline
        Value function coefficient & $c_1$ & 0.1 & 0.01\\
        \hline
        Entropy coefficient & $c_2$ & 0.001 & 0.1 \\
        \hline
        Epochs & $N$ & 30 & 10 \\
        \hline
        Time steps & $T$ & 512 & 512\\
        \hline
    \end{tabular}
    \label{tab:ppo parameters}
    \vspace{-1em}
\end{table}

\vspace{-1em}
\subsection{DS-PPO Performance}
\begin{figure}[!t]
    \centering
    \includegraphics[width=\linewidth]{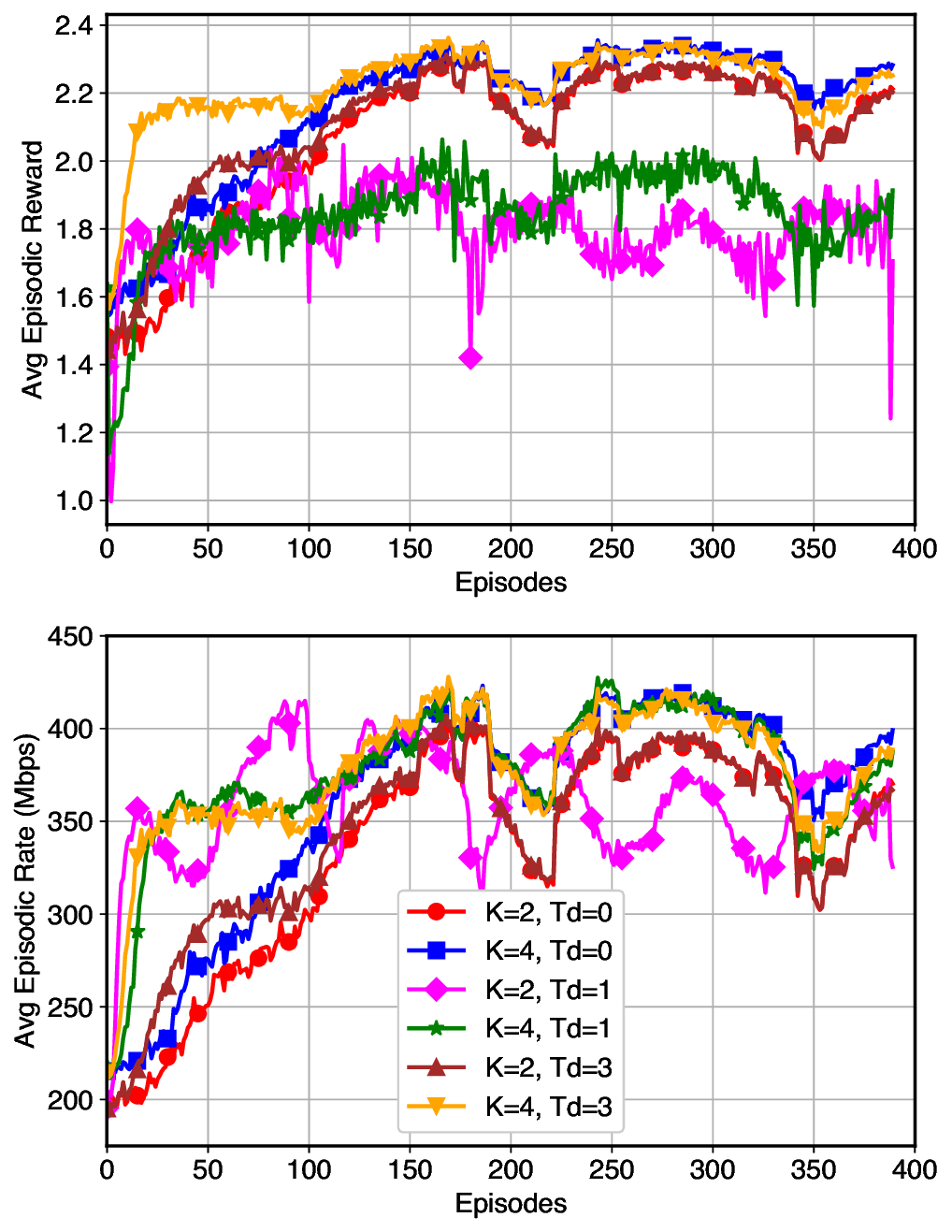}
    \caption{DS-PPO performance with $K\in\{2,4\}$ users in perfect CSI and delayed CSI , in presence of handovers. \textcolor{black}{The delayed observations effects are neglectable in both cases.}}
    \label{fig:L4_dppo_performance:handovers}
    
\end{figure}

In order to understand and evaluate the performance of DS-PPO, we have run experiments in four different settings. Specifically, we evaluate the performance of the DS-PPO in a system of $L=4$ satellites supporting $K\in\{2,4\}$ users and in the context of perfect CSI with no observation delays ($T_d =0$) and delayed CSI ($T_d =1$ and $T_d = 3$). Fig. \ref{fig:L4_dppo_performance:handovers} shows the average episodic rate and demonstrates that the episodic rate in the perfect CSI scenario is that for the delayed CSI. However, the gap between the perfect and delayed CSI scenarios is negligible, which shows the robustness of the algorithm to the delays in CSI observations. \textcolor{black}{In addition, for all the scenarios, the algorithm provides a minimum guaranteed sum rate of 300Mbps just after the 100th episode.} The effect of handovers is also shown in this figure, represented by the periodic dips in performance. The handover effect on throughput is studied in \cite{OurPaperTCOM} in more details. 
Note that as more users are served, the inter-user interference increases, leading to a reduction in the rate per user. However, this is offset by an overall increase in the sum-rate of all users. 
In the algorithmic design, though, there are significant implications that have to be considered, such as the size of the action space, which grows by 2M with the addition of each user. 

\begin{figure}[t]
    \centering
    \includegraphics[width=\linewidth]{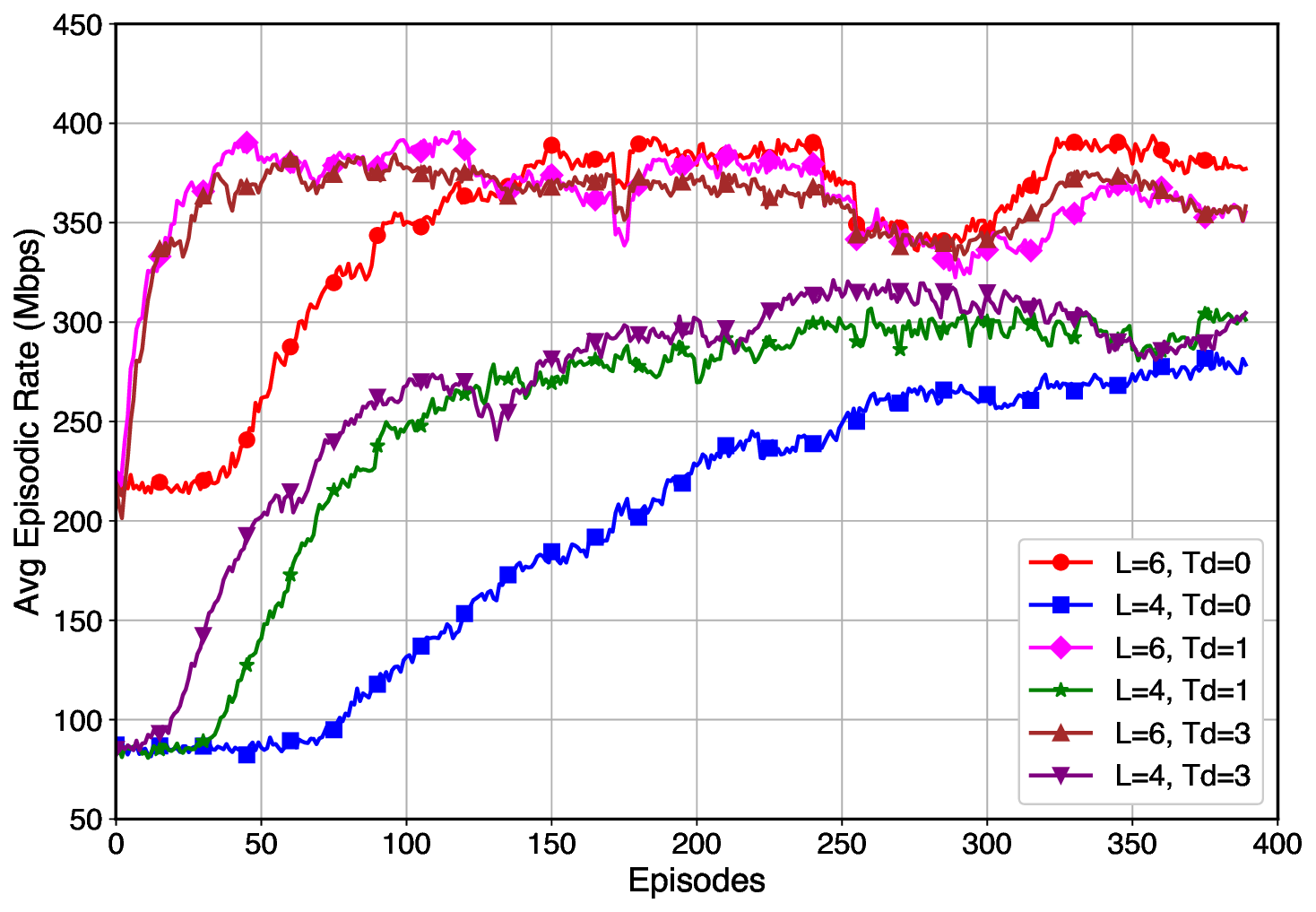} 
    \caption{Comparative results with $L\in\{4,6\}$ satellites and $K=6$ users in perfect CSI and delayed CSI scenarios.\textcolor{black}{The increment in satellites clearly increases the sum rate by $20\%$}}
    \label{fig:L4_Vs_L6_results}
\end{figure}

\begin{figure}[t]
    \vspace{-1em}
    \centering
    \includegraphics[width=\linewidth]{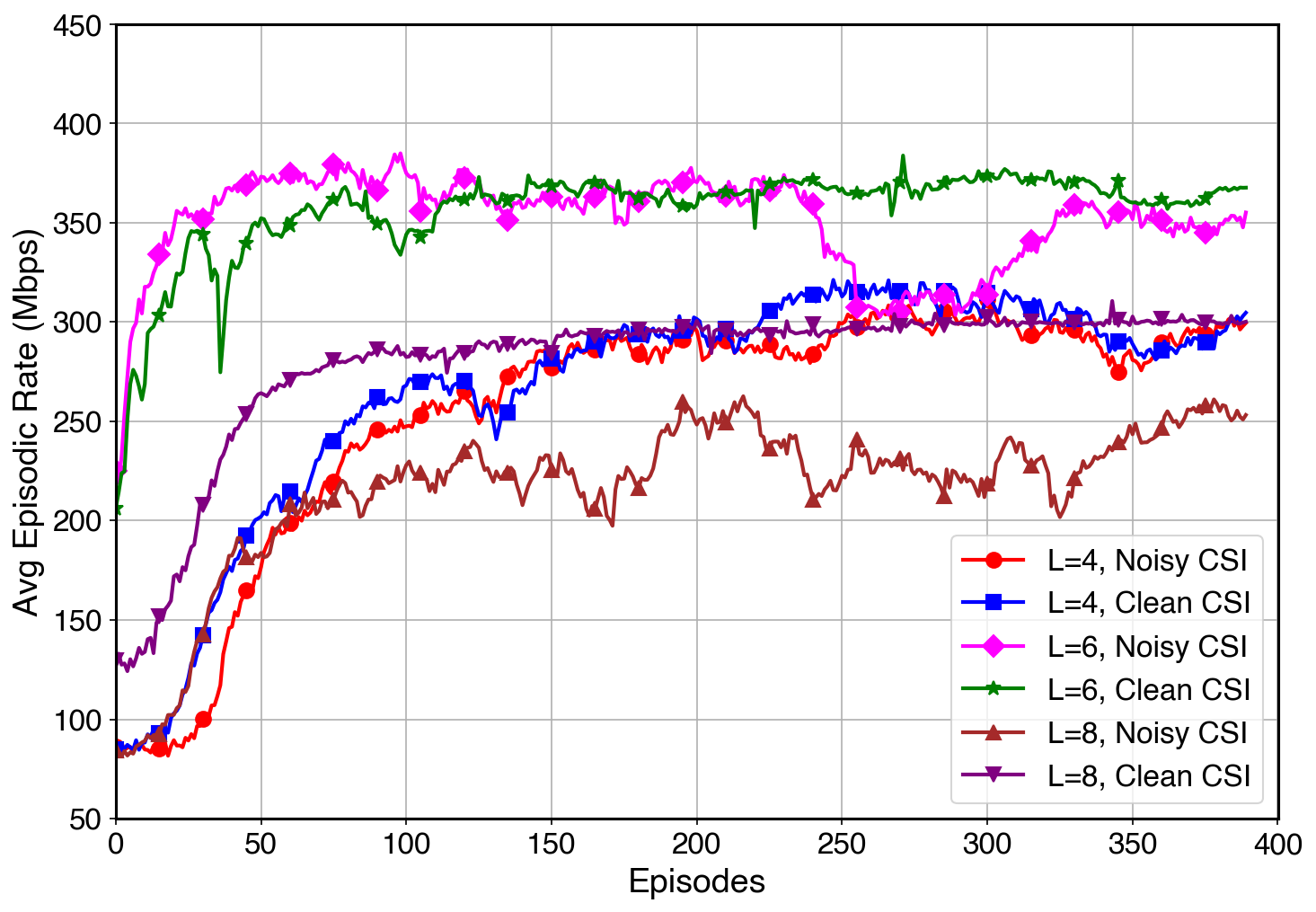}
    \caption{Comparative results with $L\in\{4,6,8\}$ satellites and $K=6$ users in delayed CSI scenarios. \textcolor{black}{The highest sum rate is guaranteed with $L=6$}}
    \label{fig:L2345678_td0}
    \vspace{-1em}
\end{figure}

\vspace{-1.25em}
\subsection{DS-PPO with varying amount of satellites}
Another important study constructed for multi-satellite systems is understanding the impacts of increasing the number of satellites on the sum-rate. In Fig. \ref{fig:L4_Vs_L6_results}, we characterize the performance of DS-PPO in perfect CSI ($T_d = 0$) and delayed CSI ($T_d = 1, \ 3$) scenarios along with $L\in\{4,6\}$ satellites. It is evident that in both the scenarios of perfect and delayed CSI, the additional satellites have helped increase the average episodic sum rate due to the increased diversity. \textcolor{black}{In addition, in the scenarios of  $L\in\{4,6\}$ satellites, the delayed CSI scenarios have achieved a similar average episodic sum rate as in the case of perfect CSI, recording evidence that DS-PPO can cope with the delayed environment. In the case of perfect CSI with $L=4$ satellites, the agent displays shallow learning, resulting in a degraded performance, but, at the convergence stage we can see that all three cases achieve almost the same sum rate.
This phenomenon is evidently associated with the PPO's sensitivity to the hyperparameters \cite{shengyi2022the37implementation}, meaning that the second stage might require tuning the hyperparameters to achieve better results.}  

\textcolor{black}{\subsection{DS-PPO's empirical degree of freedom}
In this section, we want to understand where DS-PPO achieves its maximum performance and identify what improvements are needed to further enhance its effectiveness. In Fig. \ref{fig:L2345678_td0}, we evaluate the performance of DS-PPO in the delayed CSI ($T_d = 3$) scenario in a system with $L\ = 4 ,\ 6,\ 8$ satellites and $K=6$ users. 
The sum rate is generally expected to increase by having more satellites in the system, due to the added diversity. This is shown when we increase $L=4$ to $L=6$. However, as shown in this figure, we see a 25\% fall in the sum rate when we increase the number of satellites to $L=8$. This is as a result of the added complexity to the agents. As we increase $L$, the patterns of change in the overall CSI, driven by the movements of the satellites, become more significant, making the environment increasingly non-identically distributed. Our DS-PPO algorithm is capable of handling this added complexity up to a certain point. However, beyond that, the complexity becomes too great to manage effectively. In other words, as the number of satellites grows, the optimisation problem faced by each agent becomes more challenging, which in turn impacts the performance of the second stage of PPO.}

\begin{figure}[t]
    \centering
    \includegraphics[scale = 0.55]{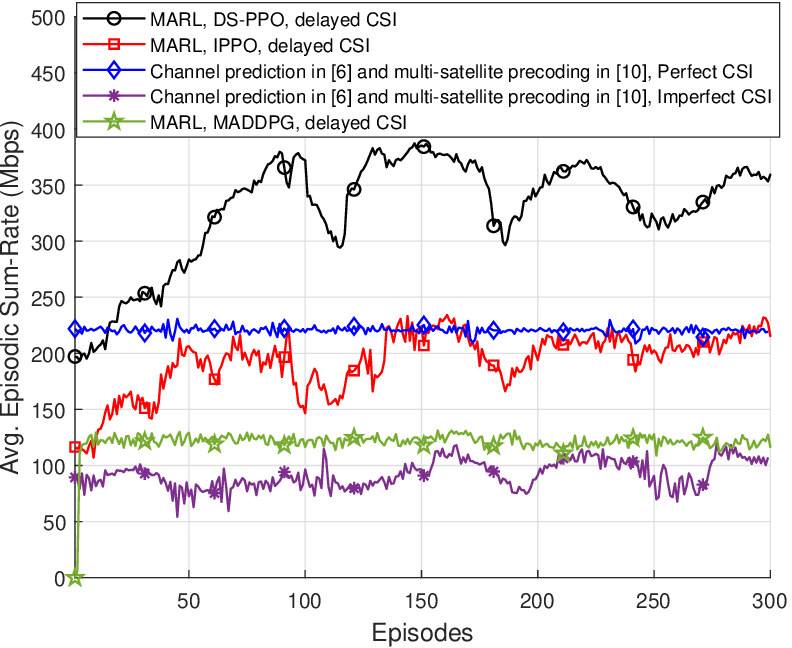}
    \caption{Comparative results with other algorithms. }
    \label{fig:L4_other_algos}
\end{figure}

\textcolor{black}{In Fig. \ref{fig:L2345678_td0}, we also evaluate the DS-PPO algorithm against realistic CSI. We assume that the delayed CSI observations are imperfect not just due to delay, but also due to other factors such as pilot contamination and estimation errors. Thus, for simulation purposes, we have assumed that the CSI observations estimated at time instant $t$ are $\hat{\mb{h}}^{[k]}(t-T_d,f)=\mb{h}^{[k]}(t-T_d,f)+\tilde{\mb{n}}_h$, $\forall k\in\{1,...,K\}$, where $\mb{\tilde{n}}_h\sim\mathcal{CN}(\mb{0},\sigma_h^2\mb{I})$ represents the channel estimation error with variance $\sigma_h^2=0.05\mathbb{E}_{t,k}\{\mb{h}^{[k]H}(t,f)\mb{h}^{[k]}(t,f)\}$.
As shown in Fig. \ref{fig:L2345678_td0}, DS-PPO maintains almost the same performance under clean and noisy CSI for $L=4$ and $L=6$, indicating that the policy learned is robust to moderate channel estimation errors in addition to CSI ageing. For $L=8$, the performance loss becomes more noticeable. This is because  the farther satellites experience weaker channels and big relative sensitivity to CSI perturbations. Nevertheless, the noisy-CSI case still preserves the main trend observed under clean CSI, confirming that DS-PPO remains effective under imperfect channel observations. Further improved robustness can be achieved by explicitly incorporating the channel estimation error statistics into the state or reward design, which is left for future work.}



\subsection{DS-PPO vs Other Algorithms}
Finally, we compare the proposed DS-PPO MARL methods, to other algorithms such as MARL with individual learning PPO (IPPO) \cite{yu2022surprising} \textcolor{black}{and MADDPG \cite{lowe2017multi}}. The agents in IPPO share only rewards between each other but no other information that influences their individual training. The algorithm for MAPPO shared information betweeen the actor and critic values, which may increase the communication overhead between the agents, and thus, we have used the IPPO which shares only the rewards. 

In Figure \ref{fig:L4_other_algos}, we show the results for a system of $L=4$ satellites and $K=4$ users. It is apparent from this figure that the DS-PPO MARL scheme achieves more than 75\% higher sum rate than the IPPO MARL, with an achieved average sum rate of 350 Mbps and variance of 15\%. This variation of the sum rate is caused by the occurrence of handovers. The results prove the expectation that MARL-IPPO will have difficulties in optimising the network due to the very large action space ($2MK = 72$) without the help of bi-level optimisation. 
Even though we use a global reward, which we expect to assist the PPO to learn and increase the sum-rate, the sum-rate is lower than the anticipated. \textcolor{\revisioncolor}{In addition, using MADDPG, we see that the resulting sum-rate is lower than that of the MARL algorithm and slighty higher that of the imperfect CSI. This is because MADDPG takes larger steps during updates, and it does not use all the updated states in buffer which results of sampling transition which are not reflected to the current state of the satellite cluster. Also, because DDPG has an actor-critic structure, the critic fails to provide meaningful information.}
The DS-PPO offers an advantage by enabling the sharing of not only the rewards but also of the singular values of the individual TPMs. This shared information allows the second stage of each agent to optimise its TPM in coordination with the individual TPMs of other agents, leading to improved overall performance.
    
In this figure, we also compare our proposed DS-PPO method—which bypasses explicit channel estimation—to other approaches that rely on CSI prediction and estimation. Specifically, if each of the $L=4$ satellites predicts its own CSI using the SatCP method \cite{9826890}, and the aggregated precoding matrix is calculated using the method of \cite{10639150}, the resulting sum rate is approximately 100 Mbps. This is only about one-third of that achieved by our DS-PPO method\textcolor{black}{For better transparency, we have summarized the performance improvements of DS-PPO in Table \ref{tab:performance_summary}.}

\begin{table}
\centering
\caption{\textcolor{black}{Performance summary of DS-PPO and baseline methods for \(L=4\), \(K=4\), and delayed CSI.}}
\label{tab:performance_summary}
\scriptsize
\setlength{\tabcolsep}{3pt}
\renewcommand{\arraystretch}{1.05}
\begin{tabular}{|p{3cm}|c|c|}
\hline
\textbf{Method} & \textbf{Avg. sum-rate } & \textbf{DS-PPO Gain} \\
\hline
DS-PPO, delayed CSI & \(\approx 350\) (Mbps) & -- \\
\hline
IPPO, delayed CSI & \(\approx 200\) (Mbps) & \(\approx 75\%\) \\
\hline
MADDPG, delayed CSI & \(\approx 125\) (Mbps) & \(\approx 180\%\) \\
\hline
CSI prediction\cite{9826890} + precoding\cite{10639150}, perfect CSI & \(\approx 220\) (Mbps) & \(\approx 59\%\) \\
\hline
CSI prediction\cite{9826890} + precoding\cite{10639150}, imperfect CSI & \(\approx 100\) (Mbps) & \(\approx 250\%\) \\
\hline
\end{tabular}
\end{table}

\section{Conclusions} \label{sec: Conclusions}
We have proposed a novel multi-agent reinforcement learning algorithm termed as DS-PPO, which aims for maximising the sum-rate of a distributed multi-satellite system.
\textcolor{black}{Our approach aims to enhance MARL algorithms to effectively handle complex environments characterized by diverse dynamic patterns, non-i.i.d. behavior, and large continuous action spaces, as commonly found in multi-satellite systems.}
Additionally, we have provided a potential solution for handling the outdated CSI observations in multi-satellite systems, where traditional methods struggle to compensate for the delay.
The numerical results demonstrate the capability of DS-PPO to achieve a high average sum-rate of 350 Mbps, exceeding the specifications of current state-of-the-art satellite systems. Furthermore, we show that DS-PPO is resilient to delayed CSI, with minor drops in performance compared to the perfect CSI scenario. In addition, we show that the computational complexity of the DS-PPO is minimal, and based on convergence analysis, \textcolor{\revisioncolor}{there is a lower bound for the performance on the second stage's action quadratically bounded on the first stage's action.}
We also evaluate the performance of DS-PPO for different numbers of users, varying numbers of satellites, in comparisons to other algorithms.
Our future work will focus on further enhancing DS-PPO to develop a more robust algorithm capable of effectively managing handovers, and extending its evaluation to other systems suffering from delayed observations. \textcolor{black}{More specific, we aim to provide a mathematical function that based on the environment, will adapt the reward in order to achieve higher downlink speeds than 300Mbps. In addition, future research will investigate advanced state-compression techniques and dynamic cluster-size selection mechanisms to reduce coordination overhead, while preserving the information required for effective distributed precoding and improved sum-rate performance.}

\section*{Acknowledgement}
The financial support of the following Engineering and Physical Sciences Research Council (EPSRC) projects is gratefully acknowledged: Platform for Driving Ultimate Connectivity (TITAN) under Grant EP/Y037243/1 and EP/X04047X/1; Robust and Reliable Quantum Computing (RoaRQ, EP/W032635/1); India-UK Intelligent Spectrum Innovation ICON UKRI-1859; PerCom (EP/X012301/1); EP/X01228X/1; EP/Y037243/1;

\bibliographystyle{IEEEtran}
\bibliography{MyRef}

\end{document}